\documentclass[10pt]{article}

\renewcommand{\textstyle}{}

\newcommand{\qed}{\vspace{.1em}\noindent\fbox{\rule{
0em}{.1em}\rule{.1em}{0em}}\vspace{1em}}

\newenvironment{proof}{

\noindent{\bf Proof:}\ }{
\hfill \qed

}

\newtheorem{theorem}{Theorem}
\newtheorem{claim}{Claim}
\newtheorem{lemma}{Lemma}

\newcommand{\OPT}{{\mbox{\sc Opt}}}
\newcommand{\LRU}{{\mbox{\sc Lru}}}
\newcommand{\FIFO}{{\mbox{\sc Fifo}}}
\newcommand{\FWF}{{\mbox{\sc Fwf}}}
\newcommand{\MARK}{{\mbox{\sc RMark}}}
\newcommand{\DMARK}{{\mbox{\sc DMark}}}
\newcommand{\PARTITION}{{\mbox{\sc Partition}}}
\newcommand{\EQUITABLE}{{\mbox{\sc Equitable}}}
\makeatletter
\newcommand{\E}{{\mathop {\operator@font E}}}
\makeatother
\newcommand{\N}{{\sf N\hspace*{-1.0ex}\rule{0.15ex}{1.3ex}\hspace*{1.0ex}}}
\newcommand{\R}{{\cal R}}
\newcommand{\RR}{{\cal RR}}

\newcommand{\newinph}{\mbox{\tt new\_in\_ph}}
\newcommand{\newbefore}{\mbox{\tt new\_bef}}
\newcommand{\new}{\mbox{\tt new}}
  
\newcommand{\prefix}{\mbox{\tt prefix}}
\newcommand{\worrisome}{\mbox{\tt worrisome}}

\begin{document}

\title{\Large On-Line Paging against Adversarially Biased Random Inputs}
\author{Neal E. Young\thanks{%
    Dartmouth College,
    Hanover NH 03755,
    ney@cs.dartmouth.edu.
    Research partially funded by NSF CAREER award CCR-9720664.
    }
  }

\date{}
\maketitle

\pagestyle{myheadings}
\markboth{}{} 
            
\begin{abstract}
  In evaluating an algorithm, worst-case analysis can be overly pessimistic.
  Average-case analysis can be overly optimistic.  An intermediate approach is
  to show that an algorithm does well on a {\em broad class} of input
  distributions.  Koutsoupias and Papadimitriou \cite{KP94} recently analyzed
  the least-recently-used (\LRU) paging strategy in this manner, analyzing its
  performance on an input sequence generated by a so-called {\em diffuse}
  adversary --- one that must choose each request probabilitistically so that no
  page is chosen with probability more than some fixed $\epsilon>0$.  They
  showed that \LRU{} achieves the optimal competitive ratio (for
  deterministic on-line algorithms), but they didn't determine the actual ratio.
  
  In this paper we estimate the optimal ratios within roughly a factor of two
  for both deterministic strategies (e.g.\ least-recently-used and
  first-in-first-out) and randomized strategies.  Around the threshold
  $\epsilon\approx 1/k$ (where $k$ is the cache size), the optimal ratios are
  both $\Theta(\ln k)$.  Below the threshold the ratios tend rapidly to $O(1)$.
  Above the threshold the ratio is unchanged for randomized strategies but
  tends rapidly to $\Theta(k)$ for deterministic ones.

  We also give an alternate proof of the optimality of $\LRU$.
\end{abstract}

\section{Introduction and Background}
The {\em paging problem} was originally studied in the context of two-level
virtual memory systems composed of a large, slow-access memory augmented with a
cache (a small, fast-access memory, holding likely-to-be accessed pages in
order to minimize access time).  

This paper concerns the following standard abstraction of this simple and
common problem.  The input is an integer $k$ and a finite sequence
$s=s_1s_2\ldots s_n$ of {\em requests}. The parameter $k$ is called {\em the
  cache size}.  The output is a {\em schedule} --- a sequence $S_1S_2\ldots
S_n$ of sets, where each set is of size at most $k$, and each $S_t$ contains
$s_t$.  Each request $s_t$ is said to occur {\em at time $t$}.  The items in $S_t$
are said to be {\em in the cache} after time $t$ up to and including time
$t+1$.  An item is said to be {\em evicted at time $t$} if the item is in
$S_{t-1}$ but not in $S_{t}$.  The {\em cost} of the schedule is number of
evictions.  A schedule for an input is {\em optimal} if it achieves the minimum
possible cost.

Next we define the paging algorithms considered in this paper.  Each evicts
pages only when the cache is full and does not contain the requested item.
{\em Least-recently-used} (\LRU) evicts the item whose most recent request is
the least recent among all items in the cache.  {\em First-in-first-out}
(\FIFO) evicts the item that has been in the cache the longest.  {\em
  Flush-when-full} (\FWF) evicts all items in the cache.  The {\em randomized
  marking algorithm} (\MARK{} \cite{FiatKLMSY91}) operates as follows.  After
an item is requested, it is marked.  When an item must be evicted, a non-marked
item is chosen uniformly at random, with the caveat that if all items in the
cache are marked, then all marks are first erased.  By a {\em deterministic
  marking algorithm}, we mean any deterministic algorithm that maintains marks
as \MARK{} does, and evicts only unmarked items.  \LRU, \FIFO, and \FWF{} are
examples.  By a {\em lazy deterministic marking algorithm} (\DMARK), we mean a
deterministic marking algorithm that evicts an item only when necessary, and
then only one item.  This additional requirement excludes \FWF.

An algorithm for the problem is {\em on-line} if, for any request sequence and
any request in that sequence, the items in the cache after the request are
independent of later requests.  In many contexts, on-line algorithms are
necessary, but on-line algorithms are necessarily sub-optimal on some request
sequences.  Hence, a natural question is how on-line algorithms can be
effectively analyzed and compared.

This paper is concerned with a generalization of the standard {\em
  competitive analysis} \cite{ST-85-2} of on-line algorithms.  The
standard model measures the quality of an algorithm $A$ by its {\em
  competitive ratio}: the minimum (to be precise, infimum) $c$ such that, for some constant
$b$, for all request sequences $s$,
$$A(s) \le c \cdot \OPT(s) + b.$$
Here $A(s)$ denotes the cost of the schedule produced by $A$ on input
$s$; $\OPT(s)$ denotes the cost of an optimal schedule.  If $A$ is a
randomized algorithm, then $A(s)$ denotes the expected cost of $A$ on
input $s$.  Note that $k$ is an implicit, and fixed, parameter in
these definitions.  Standard competitive analysis is a worst-case type
of analysis, in contrast to much of the earlier work on paging, which
is concerned with average-case analysis.\footnote{
At least one work \cite{franaszek74} preceding competitive analysis
blends average-case and worst-case analysis.
It considers input sequences where each request
is chosen from a fixed but unknown distribution on the pages,
and compares known paging strategies to the optimal {\em on-line} strategy
for that distribution.}

In the standard competitive-analysis framework the following results are known.  Any
deterministic marking algorithm, including $\LRU$, $\FIFO$, and
$\FWF$, has a competitive ratio of $k$; the ratio $k$ is the best
possible for any deterministic on-line strategy \cite{ST-85-2,BE98}.
The randomized marking algorithm \MARK{} has a competitive ratio of $2H(k)-1$
\cite{FiatKLMSY91,AchlioptasCN1996}, where $H(k)\doteq\sum_1^k 1/i \approx
\ln(1+k)$.  \PARTITION{} \cite{MS91} and \EQUITABLE{} \cite{AchlioptasCN1996},
more complicated randomized algorithms, each have competitive ratio $H(k)$.  No
randomized strategy can have a better ratio than $H(k)$ \cite{FiatKLMSY91}.

Largely due to the unrealistic magnitude of the optimal competitive ratios
\cite{young:loose-dual}, many variations on the standard model have been
considered (e.g.\ 
\cite{BorodinIRS95,FiatK95,IraniKP1996,FiatR1997,KarlinPR92,LundPR94,young:loose-dual,Young98FC}).
For a survey on competitive analysis of paging, we refer the reader to the
recent book by Borodin and El-Yaniv \cite[ch's~3-5]{BE98}.

This paper concerns the following generalization of the standard
model, recently proposed by Koutsoupias and Papadimitriou \cite{KP94}.
For any class $\Delta$ of distributions on the input sequences and
any deterministic or randomized algorithm $A$, define
$\R(\Delta,A)$, the {\em competitive ratio of $A$ against the $\Delta$-diffuse
  adversary}, to be the minimum (again, to be precise, infimum)
$c$ such that for each
distribution $D$ in $\Delta$, there is a constant $b$ such that
$$\E_D[A(r)] \le c\cdot \E_D[\OPT(r)] + b.$$
Here $r$ is a random sequence chosen according to $D$.
Define {\em the optimal ratio for deterministic on-line algorithms
  (against the $\Delta$-diffuse adversary)} to be
$$\R(\Delta) \doteq \inf_A \R(\Delta,A),$$
where $A$ ranges over all deterministic on-line algorithms.
Analogously, define {\em the optimal ratio for randomized on-line
  algorithms (against the $\Delta$-diffuse adversary)} to be
$$\RR(\Delta) \doteq \inf_{A_R} \R(\Delta,A_R)$$
where $A_R$ ranges over all randomized on-line algorithms.

The particular class of distributions considered by Koutsoupias and
Papadimitriou is denoted $\Delta_\epsilon$ and is defined as follows.  Any
distribution $D$ specifies, for each item $x$ and sequence of requests $s$, the
probability $\Pr_D(x|s)$ that the next request of the random sequence
$r$ is $x$ given that the sequence so far is $s$.  Then $\Delta_\epsilon$
contains those distributions $D$ such that, for any request sequence $s$ and
item $x$, $\Pr_D(x|s) \le \epsilon$.  The parameter $\epsilon$ is a measure of
the inherent uncertainty of each request.  Koutsoupias and Papadimitriou show
that $\LRU$ achieves the optimal ratio in this model (i.e.
$\R(\Delta_\epsilon,\LRU) = \R(\Delta_\epsilon)$), but they leave open the
question of what the ratio is.

Here we estimate the optimal ratios within roughly a factor of two,
for both deterministic and randomized algorithms.
Here is our main theorem.
\begin{theorem}
  \label{mainthm}
Define
$$\Phi(\epsilon,k) \doteq 1+\sum_{i=1}^{k-1}
{\max\{\epsilon^{-1}-i,1\}}^{-1}.$$ 
  For any $\epsilon$, let $\epsilon' = 1/\lceil \epsilon^{-1}\rceil$.
  The competitive ratios of deterministic ($\R$) and randomized ($\RR$)
  on-line algorithms against the $\Delta_\epsilon$-diffuse adversary
  are bounded as follows:
\begin{center}
  \vspace{-0.25in}
  \renewcommand{\arraystretch}{2}
  \begin{tabular}{|c||c|c||c|c|}
    \multicolumn{1}{c}{} & \multicolumn{2}{c}{%
      \raisebox{-0.15in}[0in][0in]{\underline{deterministic
          --- $\R(\Delta_\epsilon)$}}} &
    \multicolumn{2}{c}{%
      \raisebox{-0.15in}[0in][0in]{\underline{randomized
          --- $\RR(\Delta_\epsilon)$}}}\\
    \multicolumn{1}{c}{range} &
    \multicolumn{1}{c}{lower bound} &
    \multicolumn{1}{c}{upper bound} &
    \multicolumn{1}{c}{lower bound} &
    \multicolumn{1}{c}{upper bounds} \\ \hline
    $\epsilon \le 1/(k+1)$
    &$\Phi(\epsilon,k)-1$
    &$2\Phi(\epsilon,k)$ 
    &$\Phi(\epsilon',k)-1$
    &$2\Phi(\epsilon,k)$ 
    \\ \hline
    $\epsilon \ge 1/(k+1)$
    &$\Phi(\epsilon,k)$
    &$2\Phi(\epsilon,k)$ 
    &$H(k)$
    &$H(k)$
    \\ \hline
  \end{tabular}
\end{center}
The upper bound $2\Phi$ for deterministic algorithms holds for any
lazy marking algorithm (e.g.\ \LRU, \FIFO) but not for \FWF.
The upper bound $H(k)$ for randomized algorithms holds for \PARTITION{} and \EQUITABLE.
The weaker upper bound $2H(k)-1$ holds for \MARK.
\end{theorem}

In all cases except one, the competitive ratios of (lazy) deterministic and
randomized marking algorithms are at least $\Phi-1$ and at most
$2\Phi$.  The exception is that for $\epsilon$ above the threshold
$1/(k+1)$, the randomized ratio is $H(k)$ (independently of
$\epsilon$).  To
understand the behavior of the function $\Phi$, consider the case
$\epsilon = 1/n$ for some integer $n$.  Then
$$\Phi(1/n,k) = 1+H(n-1)+\cases{%
  \rule[0ex]{0em}{2ex} {-\!H(n-k)} & when $n \ge k$,
  \cr
  \rule[-1ex]{0em}{2ex} \hspace{0.5em}k-n & when $n \le k$.
  }$$
Recall that $H(k)\doteq\sum_1^k 1/i \approx \ln(k+1)$.  The threshold
of $\Phi$ around $\epsilon \approx 1/k$ is very sharp:
$$
\Phi(\epsilon,k) \textrm{ is } \cases{
  \,\le\   1 + \ln\frac{1}{\delta} & when $\epsilon = (1-\delta)/k$, \cr
  \,\approx\  \ln k & when $\epsilon = 1/k$, \cr
  \,\ge\  k\frac{\delta}{1+\delta} 
  & when $\epsilon = (1+\delta)/k$.}
$$

\section{Technical Overview}
\label{worrisome}
\label{background}
We refine an existing {\em worst-case} competitive analysis for paging
\cite{ST-85-2,FiatKLMSY91,BE98} to take into account the probabilistic
restrictions on the adversary.  We call this particular analysis the {\em
  factor-two-analysis} because for our purposes (and when used to analyze the
randomized marking algorithm \cite{FiatKLMSY91}) it (at best) can approximate
$\OPT$ only within a factor of two.

\subsection{Review of Factor-Two-Analysis in the Standard Model}
Let $A$ be any paging algorithm and let $s=s_1 s_2 \ldots s_n$ be any sequence of
requests.  The {\em phases} of $s$ partition the times $\{ 1,2,\ldots,n \}$
into intervals as follows.  Define $t(1) = 1$.  For $\ell\in\N$ inductively
define
\[t({\ell+1}) \doteq 1+\max\{ j \le n : \left|\{s_{t(\ell)},s_{t(\ell)+1},s_{t(\ell)+2}, \ldots, s_j\}\right| \le k\}.\]
For each $\ell\in\N$ such that $t(\ell)\le n$,
the {\em $\ell$th phase of $s$}
is defined to be the time interval $\{t(\ell),t(\ell)+1,\ldots, t(\ell+1)-1\}$.
Thus, during each phase except the last, $k$ distinct items are requested.

In the context of a particular time $t$, the {\em current request} refers to
the request $s_t$.  {\em This phase} or, synonymously, {\em the current phase}
means the phase containing the time $t$.  An item is {\em requested previously
  in this phase} if it is requested during this phase before time $t$.  In the
additional context of a particular schedule $S=S_1 S_2 \ldots S_n$ for $s$,
{\em the cache} refers to the set $S_{t-1}$ of items in the cache before
request $t$.  Then at each time $t$, each item is classified with respect to its
status before request $s_t$ as follows:
\begin{description}
\item{\bf new} --- not requested previously in this phase or in the last phase.
\item{\bf old} --- requested during the last phase, but not previously in this phase.
\item{\bf redundant} --- requested previously in this phase.
\item{\bf worrisome} --- requested in the last phase or previously in this phase, but not
  in the on-line algorithm's cache.
\end{description}
Each {\em request} is classified as well, according to the status of the
requested item.  For instance, a request $s_t$ is {\em new} if the requested
item was new after request $s_{t-1}$.  Each phase (except possibly the last)
has $k$ non-redundant requests, each one of which is either new or old.
Define
\begin{description}
\item{$\new(s)$} --- the total number of new requests in sequence $s$.
\item{$\newinph(\ell)$} --- (in the context of some sequence) the
  total number of new requests in the $\ell$th phase of the sequence.
  Here $\ell$ is any positive integer.  If $\ell=0$ or there is no
  $\ell$th phase, define $\newinph(\ell)$ to be $0$.
\end{description}
The relevance of the new requests is as follows.
\begin{lemma}[\cite{FiatKLMSY91,youn-91-2}]\label{newlemma}
$\new(s)/2 \le \OPT(s) \le \new(s)$
\end{lemma}
\begin{proof}
  Consider the $(\ell-1)$st and $\ell$th phases of $s$ for any $\ell$.
  The number of distinct items requested in the two phases is
  $k+\newinph(\ell)$.  Thus, the number of evictions incurred by
  \OPT{} during the two phases is at least $\newinph(\ell)$ and
  \begin{eqnarray*}
    \OPT(s) &\ge& \max\bigg\{
    \sum_{\ell \mbox{ \footnotesize odd}} \newinph(\ell),
    \sum_{\ell \mbox{ \footnotesize even}} \newinph(\ell)\bigg\} \\
    &\ge& \sum_\ell \newinph(\ell)/2 = \new(s)/2.
  \end{eqnarray*}
  
  On the other hand, the following schedule costs at most $\new(s)$.
  At the beginning of each phase, evict those items that are not
  requested during the phase and bring in the items that are not in
  the cache but are requested during the phase.  After each phase
  ends, the items requested during that phase are in the cache, so the
  number of evictions in the next phase is just the number of new
  requests in that phase.  Thus, the cost of this schedule is
  $\new(s)$.  Since the schedule produced by \OPT{} is at least as
  good, $\OPT(s) \le \new(s)$.
\end{proof}

By the {\em amortized} cost incurred by \OPT{} during a phase, we mean half the
number of new requests in that phase.  By the lemma above, the total cost
incurred by \OPT{} is at least the total of these amortized costs and at most
twice the total.  To show bounds on the competitive ratio of $A$, we use
the standard method of bounding the cost incurred by $A$ during a phase divided
by the amortized cost incurred by $\OPT$ during the phase.  For instance, if
this ratio is at most $c$ for each phase of a sequence $s$, then it follows
immediately that $A(s) \le c\cdot\OPT(s)$.

One intuition for understanding \MARK{} and other marking algorithms
such as \LRU{}, \FIFO{}, and even \FWF{} is that they are emulating
the schedule described in the proof above that $\OPT(s)\le \new(s)$.
That is, during each phase, the ``goal'' (intuitively speaking) is to
get the items that will be requested during the phase into the cache.
From this point of view, once an item is requested during a phase, it
should be kept in the cache.  This is the principle that defines a
deterministic marking algorithm.

If this principle is followed, then only non-redundant requests can
cause evictions.  Since the phase ends after $k$ non-redundant requests, any
deterministic marking algorithm incurs a cost of at most $k$ during
the phase.  This means that in the standard model, the competitive
ratio is at most $k$ ($\OPT$ also incurs at least one eviction per
phase).  Conversely, the adversary can force a ratio of $k$ against a
deterministic on-line algorithm by making one new request each phase
and then making $k-1$ requests, each to whichever old item is not
currently in the cache.

\subsection{Factor-Two-Analysis for the Diffuse Adversary}
In the standard model, the adversary can force each old request (i.e.,
each non-redundant request to an item requested in the previous phase)
to cause an eviction.  In the diffuse adversary model, this is not so,
because the adversary can only assign $\epsilon$ probability to each
item.  The adversary may have to assign probability to redundant
and/or new items.  To adapt the standard analysis to the diffuse
adversary setting, we analyze the extent to which the adversary can
assign probability to old items.  Recall that old items that are not
in the on-line algorithm's cache are called {\em worrisome}, as are
requests to such items.  We analyze the extent to which the adversary
can cause worrisome requests.

We next sketch the argument for the upper bound, glossing over issues
of probabilistic conditioning, in order to convey the intuition.  In
the subsequent section we give a formally correct treatment.  We then
give the lower bound; the intuition for the lower bound is similar to
that of the upper bound.

Consider the $\ell$th phase for any $\ell$.  There are $k$
non-redundant requests in the phase (except possibly for the last
phase, which may have fewer).  Consider the state of any marking
algorithm \DMARK{} just before the $(i+1)$st non-redundant request,
for $1\le i\le k-1$.

The $i$ redundant items are marked and in the cache.
Of the $k$ items requested last phase, at most $\newinph(\ell)$ are
worrisome (out of the cache).  Thus, the adversary can assign at
most $\epsilon\, \newinph(\ell)$ probability to worrisome items.
Since there are only $i$ redundant items, the adversary has to assign
at least $1-\epsilon i$ probability to non-redundant items.
Therefore, the probability that the request will be worrisome,
given that the request turns out to be non-redundant, is at most
$$\frac{\epsilon\, \newinph(\ell)}{1-\epsilon i} =
\frac{\newinph(\ell)}{\epsilon^{-1}- i}$$
(or 1 if this quantity is negative or more than 1).  Summing over $i$,
adding $\newinph(\ell)$ for the evictions due to new requests, and
dividing by $\newinph(\ell)/2$ (the amortized cost incurred by \OPT{}
for the phase) gives the desired upper bound $2\Phi$ on the competitive
ratio.

The above upper bound can be turned into a roughly equivalent lower
bound.  The lower bound loses a factor of 2 because of our use of new
requests in approximating $\OPT(s)$.  It loses an additional additive
term of 1 in some cases; we revisit this issue after presenting the
lower bound.

\section{Upper Bound for Deterministic Algorithms}
\label{sec:dub}
Next we prove the upper bounds on deterministic strategies in Theorem~\ref{mainthm}:
\begin{lemma}
\label{ublemma}
For any lazy deterministic marking algorithm \DMARK{}
and $D\in\Delta_\epsilon$,
$$\E_D[\DMARK(r)] \le 2\Phi(\epsilon,k)\cdot \E_D[\OPT(r)] + O(1)$$
\end{lemma}
\begin{proof}
  Without loss of generality, assume that $D$ generates only sequences whose last
  phase has $k$ non-redundant requests.  (Otherwise we can easily
  modify the distribution so that the condition is satisfied, while
  increasing $\E[\OPT(r)]$ by at most the constant $k$.)
  In the context of the random sequence $r$,
  define the following random variables and events.
  \begin{description}
  \item[$R_{\ell,i}$] --- the $(i+1)$st non-redundant request in the
    $\ell$th phase of $r$, if there is an $\ell$th phase.
  \item[$\prefix(R)$] --- the prefix of $r$ up to but not including
    request $R$ of $r$.  
  \item[$\newbefore(R)$] --- the number of new requests before request $R$
    in the phase of $r$ containing $R$.
  \item[$\newinph(\ell)$] --- the total number of new requests made
    in the $\ell$th phase of $r$, if there is an $\ell$th phase,
    otherwise $0$.
  \item[$\worrisome(R)$] --- the event that request $R$ of $r$ is worrisome.  
  \end{description}
  In what follows, we abuse notation slightly as follows.  By the
  event ``$\prefix(R_{\ell,i})=s$'', we mean ``there is an $\ell$th
  phase in $r$ and the prefix of $r$ preceding request $R_{\ell,i}$
  is sequence $s$''.  Similarly, by the event
  ``$\worrisome(R_{\ell,i})$'', we mean ``there is an $\ell$th phase
  in $r$ and the request $R_{\ell,i}$ in that phase is worrisome''.

  We start by proving the following claim:
  \begin{claim}\label{condclaim}
    Fix any $\ell$ and $i$ $(1 \le i \le k-1)$.
    Let $s$ be any sequence such that the event
    $\prefix(R_{\ell,i})=s$ can happen.
    That is, $s$ has $\ell$ phases, and the last phase of $s$ has $i$
    non-redundant requests.
    Then
    $$\Pr[\worrisome(R_{\ell,i}) ~|~ \prefix(R_{\ell,i}) = s]
    \le \E\bigg[\frac{\newbefore(R_{\ell,i})}{\max\{1,\epsilon^{-1}-i\}}~
    \bigg|~ \prefix(R_{\ell,i})=s\bigg].$$
  \end{claim}
  Conditioning on ``$\prefix(R_{\ell,i})=s$'' lets us use the restrictions on the adversary.
  
  Here is the proof of Claim~\ref{condclaim}.  In the event that $s$ is a
  prefix of $r$, consider the random variable $r_t$ where $t=|s|+1$.
  (There must be such a request because $i<k$ and each phase of $r$,
  including the last, by the assumption at the beginning of the proof, has $k$
  non-redundant requests.)

  The event $\prefix(R_{\ell,i})=s$ happens if and only if $s$ is a prefix of
  $r$ and $r_t$ is non-redundant.  If $\prefix(R_{\ell,i})=s$, then the event
  $\worrisome(R_{\ell,i})$ happens if and only if $r_t$ is worrisome.  Thus,
  \begin{eqnarray*}
    \lefteqn{\Pr(\worrisome(R_{\ell,i}) ~|~ \prefix(R_{\ell,i}) = s)}
    \\& = & \Pr(\worrisome(r_t)~|~
    s \mbox{ is a prefix of $r$ and $r_t$ is non-redundant})
    \\& = & \frac{\Pr(\worrisome(r_t) ~|~ s \mbox{ is a prefix of } r)}{%
      \Pr(r_t \mbox{ is non-redundant $|~ s$ is a prefix of } r)}.
  \end{eqnarray*}
  Assume that $s$ is a prefix of $r$.  After processing $s$, $\DMARK$
  has all but $\newbefore(r_t)$ of the items requested in the previous
  phase in the cache.  Thus, the adversary can assign at most
  $\epsilon\,\newbefore(r_t)$ probability to worrisome items.  Thus, the
  numerator above is at most $\epsilon\,\newbefore(r_t)$.  Since there have
  been $i$ non-redundant requests in this phase before $r_t$, there are
  only $i$ redundant items, so the denominator above is at least
  $1-\epsilon i$.
  To finish the proof of Claim~\ref{condclaim}, note that
  $\E[\newbefore(R_{\ell,i})~|~\prefix(R_{\ell,i})=s] = \newbefore(r_t)$.

  Now fix $i$ and $\ell$.  In the set of events
  $\{\prefix(R_{\ell,i})=s ~|~ s \mbox{ is a sequence}\}$,
  exactly one event happens.
  Thus, the bound in Claim~\ref{condclaim} holds unconditionally:
  $$\Pr[\worrisome(R_{\ell,i})]
  \le \E\bigg[\frac{\newbefore(R_{\ell,i})}{\max\{1,\epsilon^{-1}-i\}}\bigg].$$
  Since $\newbefore(R_{\ell,i}) \le \newinph(\ell)$, 
  it follows that for all $\ell$ and $i$,
  $$\Pr[\worrisome(R_{\ell,i})]
  \le \E\bigg[\frac{\newinph(\ell)}{\max\{1,\epsilon^{-1}-i\}}\bigg].$$
  Since $\DMARK(r)$ is the number of new or worrisome requests in $r$,
  \begin{eqnarray}
    \E[\DMARK(r)] & \le & \textstyle
      \E\Big[\sum_\ell \newinph(\ell)
        + \sum_{\ell,i} {\frac{\newinph(\ell)}{\max\{1,\epsilon^{-1}-
            i\}}}\Big]
    \\ & = & \textstyle \big(1+\sum_i \max\{1,\epsilon^{-1}- i\}^{-1}\big)
    \,\cdot\,\E\Big[\sum_\ell \newinph(\ell)\Big]
    \\ & = & \Phi(\epsilon,k) \; \E[\new(r)]
    \\ & \le & \Phi(\epsilon,k) \; \E[\OPT(r)/2] ~~~\mathrm{(by~Lemma~\ref{newlemma})}.
  \end{eqnarray}
\end{proof}

\section{Lower Bound for Deterministic Algorithms}
Next we prove the lower bounds on deterministic strategies in Theorem~\ref{mainthm}:
\label{sec:dlb}
\begin{lemma}
  \label{lblemma}
  For any $\epsilon>0$, any $k$, and any deterministic on-line
  algorithm $A$, there is a distribution $D\in\Delta_\epsilon$ such
  that
  $$\E_D[A(r)] \ge (\Phi(\epsilon,k) -1+1/m)\cdot\E_D[\OPT(r)].$$
  where $m=\max\{1,\lceil\epsilon^{-1}\rceil-k\}$, 
  and $\E_D[\OPT(r)]$ is arbitrarily large.
\end{lemma}
\begin{proof}
  We describe $D$ by describing an adversary that requests items
  probabilistically subject to the limitations of $\Delta_\epsilon$.
  Fix $\epsilon>0$ and $k>0$.  Assume $\epsilon>1/2k$ (otherwise the
  desired lower bound is trivially satisfied, because $\Phi(1/2k,k)-1+1/m$
  is less than $1$).  
  
  The adversary requests the items in an on-line fashion, phase by
  phase.  In the first part of each phase, the adversary makes $m$ new
  requests by assigning probability only to items not previously
  requested.
  
  For each remaining request, the adversary assigns a probability to
  each item as follows.  First priority is given to worrisome items
  (those previously requested in this phase or in the last one but not
  in the cache of $A$).  Second priority is given to redundant items
  (those requested previously in this phase and in the cache).  Third
  priority is given to the remaining old items (the items not yet
  requested this phase, but in the cache).
  
  Items are selected in order of priority and assigned as much
  probability as possible, subject to the constraint that no item is
  assigned probability more than $\epsilon$ and the total probability
  assigned is $1$.  By the choice of $m$, we have $(k+m)\epsilon \ge
  1$, so all three kinds of items suffice for all probability to be
  assigned.
  
  The adversary follows this strategy until $k$ distinct items have
  been requested, at which point the adversary begins a new phase.
  The adversary continues for $N$ phases, where $N$ is arbitrarily
  large so that $\OPT(r)$ is also arbitrarily large.
  
  This defines the distribution $D \in \Delta_\epsilon$.  Let $r$ be a
  random request sequence chosen from $D$.  Next we prove that $\E[A(r)] \ge
  Nm(\Phi(\epsilon,k) -1+1/m)$.  This proves the claimed bound, since
  $\OPT(r) \le Nm$ (by Lemma~\ref{newlemma}).  Consider any $\ell$
  s.t. $1\le \ell \le N$.  For $i=m,\ldots,k-1$, define
  \begin{description}
  \item[$\worrisome(R_{\ell,i})$] --- the event that the $i$th
    non-redundant request of the $\ell$th phase is worrisome.
  \end{description}
  The expectation of $A(r)$ is $Nm+\sum_{\ell,i} \Pr[\worrisome(R_{\ell,i})]$.
  For any $\ell$ and $i$ s.t. $m\le i\le k-1$, consider the time just
  before the $(i+1)$st non-redundant request of the $\ell$th phase.
  There have been $i$ non-redundant requests so far in the phase, so
  there are $i$ redundant items.  There have been $m$ new requests so
  far, so there are $k+m$ items that were requested last phase or
  already this phase.  Since the on-line algorithm has at least $m$ of
  these items not in the cache, there are at least $m$ worrisome
  items.  Thus, the adversary assigns at least $\epsilon m$
  probability to worrisome items and at least $\epsilon i$ probability
  to redundant items.  (Unless $\epsilon m + \epsilon i > 1$, in which
  case $\Pr[\worrisome(R_{\ell,i})]=1$ --- the adversary forces a
  worrisome request.)  Thus, the probability that the request is
  worrisome, conditioned on it being non-redundant, is 
  $$\Pr[\worrisome(R_{\ell,i})] \ge \frac{\epsilon m}{\max\{1-\epsilon
    i,\epsilon m\}} = \frac{m}{\max\{\epsilon^{-1}-i,m\}}
  = \frac{m}{\max\{\epsilon^{-1}-i,1\}}.
  $$
  The rightmost equality holds because the choice of $m$
  implies that either $m=1$ or $\epsilon^{-1}-i\ge m$.
  Adding the $m$ new requests and summing over $i=m,\ldots,k-1$, the
  expected cost to $A$ for each of the $N$ phases is at least
  $$
  m+\sum_{i=m}^{k-1} \frac{m}{\max\{\epsilon^{-1}- i,1\}} \\
  \ge
  1+\sum_{i=1}^{k-1}\frac{m}{\max\{\epsilon^{-1}- i,1\}}.$$
  The rightmost expression is $m\,(\Phi(\epsilon,k)-1+1/m)$.
\end{proof}
The adversary can probably be made a little stronger to get a slightly
better lower bound when $\epsilon \le 1/(k+1)$.  In this case the
issue of how the optimal adversary should fix $m$ appears to be
relatively subtle.  This is why the lower bound loses the additive $1$
with respect to the upper bound in this case.  One small improvement
to the above adversary would be, when the adversary is requesting new
items, to use the opportunity to also allocate probability to
worrisome items.

\section{Randomized Strategies}
\label{sec:rand}
In this section we finish the proof of Theorem~\ref{mainthm} by proving the
upper and lower bounds for randomized strategies claimed there.  By using what
we already know, very little work is required to get the bounds.

We first consider lower bounds.  Fix $\epsilon>0$ and $k>0$.  We start
with the case $\epsilon \le 1/(k+1)$.  For simplicity we make the
technical assumption that $\epsilon^{-1}$ is an integer.  This
assumption is not too restrictive and allows us to reuse the
deterministic lower bound as follows.

\begin{lemma}\label{independence}
  If $\epsilon^{-1}$ is an integer greater than $k$, then the
  distribution $D$ described in the proof of Lemma~\ref{lblemma} is
  independent of the algorithm $A$.
\end{lemma}
\begin{proof}
  Consider that distribution.  Within each phase, the random sequence
  $r$ has requests to $m$ new items, followed by requests restricted
  to a set of $k+m$ items, where $m=\max\{1,\epsilon^{-1} -k\}$, until
  $k$ distinct items have been requested.  The condition on $\epsilon$
  and the choice of $m$ imply that $m=\epsilon^{-1} -k$, so that
  $\epsilon=1/(k+m)$.  In this case, each phase simply consists of
  requests to $m$ new items, followed by a sequence of requests to the
  $k+m$ items, where each request is chosen {\em uniformly at random} from
  those $k+m$ items, until a total of $k$ distinct items have been
  requested, after which the next phase begins.
\end{proof}

This distribution generalizes a distribution defined in a previous lower bound
on the competitive ratio of randomized on-line strategies against the standard
adversary \cite[Thm.~8.7]{BE98}, \cite[Thm.~13.2]{MotwaniRa95}.  (That lower
bound is equivalent to our case $m=1$.)  There and here, Yao's principle
implies that for a random input $r$ from any input distribution $D$, any
randomized on-line algorithm $A_R$ satisfies
$$\E[A_R(r)] \ge \inf_A \E[A(r)]$$
where $A$ ranges over all
deterministic on-line algorithms.

(Briefly, this is because $A_R$ may
be viewed as probabilistically picking some deterministic algorithm
$A$, and then running $A$ on the input $r$.
Thus,
$\E_D[A_R(r)] = \sum_A \Pr[A_R \mbox{ chooses } A]\cdot \E_D[A(r)]
\ge \inf_A \E_D[A(r)]$.
Here $D$ can be any distribution, but we take it to be the one
defined in Lemma~\ref{lblemma}.  The input $r$ is randomly chosen from $D$.
We refer the reader to \cite[Thm.~13.2]{MotwaniRa95}
or \cite[Thm.~8.7]{BE98} for a full explanation of Yao's principle
in this context.)

By Lemma~\ref{independence}, in the special case when $\epsilon^{-1}$ is an
integer greater than $k$, the distribution defined in the previous section is
independent of the on-line algorithm $A$.  Thus, by Yao's principle, the lower
bounds proved there extend to randomized algorithms.  This proves:
\begin{lemma}
\label{rlblemma}
Suppose $\epsilon \le 1/(k+1)$ and $\epsilon^{-1}$ is an integer.
Then the lower bound established in Lemma~\ref{lblemma} also applies
to randomized on-line algorithms.
\end{lemma}
Decreasing $\epsilon$ only weakens the adversary.
Thus, when $\epsilon$ is not an integer,
letting $\epsilon' = 1/\lceil \epsilon^{-1}\rceil<\epsilon$,
the lower bounds hold with $\epsilon'$ replacing $\epsilon$.

Also, when $\epsilon = 1/(k+1)$ it is easy to verify that the above lemma
implies that the ratios are at least $H(k) \doteq \sum_1^k 1/i$.
This proves:
\begin{lemma}
\label{rlblemma2}
Suppose $\epsilon \ge 1/(k+1)$.
Then $\RR(\Delta_\epsilon) \ge H(k)$.
\end{lemma}
So the above two lemmas prove the lower bounds for randomized strategies claimed
in Theorem~\ref{mainthm}.
What about the upper bounds?  Because the diffuse adversary is no stronger 
than the standard adversary, we get immediately from previous results that:
\begin{lemma}
\label{rublemma}
For $\epsilon \le 1/(k+1)$,
$\RR(\Delta_\epsilon,\MARK) \le 2\Phi(\epsilon,k)$.

For $\epsilon \ge 1/(k+1)$,
$\RR(\Delta_\epsilon,\MARK) \le 2H(k)-1$,
while
$\RR(\Delta_\epsilon,\PARTITION) \le H(k)$,
and
$\RR(\Delta_\epsilon,\EQUITABLE) \le H(k)$.
\end{lemma}
The first upper bound follows from the fact that Lemma~\ref{ublemma} also
applies to $\MARK$ (since the upper bound applies to any deterministic marking
algorithm, i.e., any conditioning of $\MARK$ on a particular outcome of its
random choices).  The remaining upper bounds follow from known upper bounds
on the competitive ratios of the various algorithms
against the (stronger) standard adversary
\cite{BE98,FiatKLMSY91,MS91,AchlioptasCN1996}.
Lemma~\ref{rublemma} proves the upper bounds on randomized strategies
in Theorem~\ref{mainthm}.  This completes the proof of that theorem.

\section{Alternate Proof that $\LRU$ is Optimal}
For the record, we include here a ``distillation'' of Koutsoupias and
Papadimitriou's proof that \LRU{} is optimal against the diffuse adversary
$\Delta_\epsilon$.
This version of the proof is shorter and self-contained, but does not give the
intermediate results about work functions in the original proof.

\newcommand{\C}[1]{{\mbox{rank}({#1})}}
\newcommand{\Q}[1]{\mathsf{#1}}

Given a request sequence $s$ of items from a universe $U$, and an (arbitrary)
initial ordering $\pi$ of the items, define the {\em rank of an item $x\in U$
  in $s$} to be the rank of $x$ in the following ordering: items that are
requested in $s$ are first, in order of last request; items that are not
requested in $s$ are next, ordered by $\pi$.

In analyzing an on-line paging algorithm, if $s$ is the sequence of requests
seen so far, then the most recently requested item currently has rank 1, the
next most recently requested item currently has rank 2, etc.  Without loss of
generality, when specifying a request or the contents of the cache, we can
specify each item by its current rank; this uniquely identifies the item.
Except in the proof of Lemma~\ref{domlemma} where we use both representations,
{\em items in this section are assumed to be specified by their current rank}.

\begin{lemma}
  \label{domlemma}
  Let $r$ and $r'$ be two equal-length request sequences.  Let $\Q{r}$ and
  $\Q{r}'$, respectively, be the same sequences but with each request specified by
  rank (w.r.t.\ the same initial ordering and universe).  If $\Q{r}$ dominates
  $\Q{r}'$ in the sense that $\Q{r}_t\ge\Q{r}'_t$ for all $t$, then
  $\OPT(\Q{r})\ge\OPT(\Q{r}')$.
\end{lemma}
\begin{proof}
  \newcommand{\T}{d}
  It suffices to prove the case when there is a single $\T$ such that
  $\Q{r}'_\T = \Q{r}_\T - 1$ but $\Q{r}_t = \Q{r}'_t$ for all $t \neq \T$.  The
  general case then follows by induction.  Assume such a $\T$.
  
  How do $r$ and $r'$ differ?  Consider the two sequences simultaneously for
  $t=1,2,\ldots,|r|$ in an on-line fashion.  At each $t$ focus on the ranks of the
  items in the two subsequences $s=r_1r_2\ldots r_t$ and
  $s'=r'_1r'_2r'\ldots r'_t$.
  
  At each time $t<\T$, for each item, the rank in $s$ equals the rank in $s'$.
  Let $x$ and $x'$ be the items requested, respectively, in $r$ and $r'$ at
  time $d$.  By assumption, just before time $\T$, the respective ranks of $x$
  and $x'$ are $\Q{r}_\T$ and $\Q{r}_\T-1$.  What about just after time $\T$?
  In sequence $s$, the rank of $x$ changes to $1$, while the rank of $x'$
  changes to $\Q{r}_\T$.  In sequence $s'$, the rank of $x$ stays $\Q{r}_\T$,
  while the rank of $x'$ changes to $1$.  For each item other than $x$ or $x'$,
  the rank of the item is equal in both sequences.
  
  This means that the sequence of items requested by $r$ is the same as the
  sequence of items requested by $r'$, except that from time $\T$ to the end,
  the roles of $x$ and $x'$ are reversed: if $r$ requests $x$ (resp.~$x'$),
  then $r'$ requests $x'$ (resp.~$x$).
  
  Let $i$ and $i'$, respectively, be the times of the most recent requests to
  $x$ and $x'$ before time $\T$.  (If either item is being requested for the
  first time, then let $i$ or $i'$ equal $1$, as appropriate.)  By assumption
  $\Q{r}'_\T=\Q{r}_\T-1$, so $i\le i'$.
  
  Consider any schedule $S$ for $r$.  For any $j$ with $i'< j \le \T$, consider
  obtaining $S'$ from $S$ by reversing the roles of $x$ and $x'$ from time $j$
  onward (i.e.\ swapping the two in $S_{j},S_{j+1},\ldots$).  By the
  established relation between $r$ and $r'$, $S'$ will be a valid schedule for
  $r'$.  To finish, we need only choose $j$ so that $S'$ costs no more than
  $S$.  In particular, at time $j$, $S'$ should evict no more of the two pages
  $\{x,x'\}$ than $S$ does.  If for some $j$, $|\{x',x\} \cap S_j| \in\{0,2\}$
  or $|\{x',x\} \cap S_{j-1}| \in\{0,2\}$, then this
  $j$ clearly suffices.  Otherwise there is a $j$ such that
  $\{x',x\} \cap S_{j-1} = \{x'\}$
  and
  $\{x',x\} \cap S_{j} = \{x\}$.  Using this $j$, $S'$ is cheaper than $S$.
\end{proof}

\begin{theorem}[\cite{KP94}]
  Let $D$ be any distribution $D\in\Delta_\epsilon$.
  Let $A$ be any deterministic on-line algorithm.
  Then there is a distribution $D'\in\Delta_\epsilon$
  such that
  $$\E_D[\LRU(r)] \le \E_{D'}[A(r')] \mbox{~~and~~}
  \E_D[\OPT(r)] \ge \E_{D'}[\OPT(r')],$$
  where $r$ and $r'$ are randomly chosen according to $D$ and $D'$, respectively.

  Thus, $\R(\Delta_\epsilon) = \R(\Delta_\epsilon,\LRU)$.
\end{theorem}
\begin{proof}
  In what follows, we assume all items are specified not by name but by rank
  (with respect to some sequence implicit in context, the universe $U$ of the
  items requested by $D$, and an arbitrary initial ordering).
  
  Intuitively, the argument is the following.  At each request, we pair each
  page $\Q{x}$ in $A$'s cache but not in \LRU's cache with a unique page
  $f(\Q{x})$ in \LRU's cache but not in $A$'s.  For each such $\Q{x}$, if $D$
  assigns more probability to $\Q{x}$ than to $f(\Q{x})$, then we shift some of
  the probability from $\Q{x}$ to $f(\Q{x})$.  This gives us a modified
  assignment of probabilities to pages for the request; in this way we define
  $D'$.  We show that this shifting procedure ensures that at each request, $A$
  is as likely to fault (on a request from $D'$) as $\LRU$ was (on the
  corresponding request from $D$).  Furthermore, $D'$ is better for \OPT\ than
  $D$ is, because when we shift probability from $\Q{x}$ to $f(\Q{x})$, we know
  that, as $\Q{x}$ is not in \LRU's cache but $f(\Q{x})$ is, we are shifting
  probability from a higher-ranked page to a lower-ranked page (in the sense of
  Lemma~\ref{domlemma}).
  
  Formally, the following random experiment defines the distribution $D'$ by describing
  how to choose a random sequence $\Q{r'}$ according to that distribution.
  Choose a random sequence $\Q{r}$ according to $D$.  Reveal $\Q{r}$ in an
  on-line fashion, one request at a time, producing each corresponding request
  of $\Q{r}'$ as follows.

  Let $\Q{L}$ denote the cache of \LRU{} (specified by rank with respect to
  $\Q{s}$) after processing $\Q{s}=\Q{r}_1\ldots \Q{r}_{t-1}$.  Similarly, let
  $\Q{A}$ denote the cache (specified by rank {\em with respect to $\Q{s}'$})
  of $A$ after processing $\Q{s}'=\Q{r}'_1\ldots \Q{r}'_{t-1}$.  Let $f$ be any
  1-1 mapping from $\Q{A}-\Q{L}$ into $\Q{L}-\Q{A}$ (note $|\Q{A}|\le|\Q{L}|$)
  and define (in the context of $\Q{s}$ and $\Q{s}'$)
  \begin{eqnarray*}
    \mathcal{X} &\doteq&
    \left\{\Q{x} \in \Q{A}-\Q{L} \,\big|\, p(f(\Q{x})) < p(\Q{x})\right\}, \mbox{ where}
    \\ p(\Q{x}) & \doteq & \textstyle \Pr_D(\Q{x}|\Q{s}).
  \end{eqnarray*}
  $\mathcal{X}$ is the set of pages from which we want to shift probability.
  
  Finally, determine $\Q{r}_t$ as follows.  First set $\Q{r}_t'=\Q{r}_t$, but
  if $\Q{r}_t\in\mathcal{X}$, {\bf change} $\Q{r}_t'$ to $f(\Q{r}_t)$ with
  probability $p(f(\Q{r}_t))/p(\Q{r}_t)$.

  This completes the random experiment that gives $\Q{r'}$ and so defines
  $D'$.  Each outcome of this experiment determines a {\em pair} of
  random variables $(\Q{r},\Q{r}')$.
  
  We use ``$\Pr_{D'}(X|\Q{s},\Q{s'})$'' to denote the probability of event $X$
  conditioned on $\Q{s}$ and $\Q{s}'$ being prefixes of $\Q{r}$ and $\Q{r'}$.
  The following claim characterizes the distribution of $\Q{r}'_t$ conditioned
  on this event.
  \begin{claim}\label{swap}
    Fix any two sequences $\Q{s}$ and $\Q{s}'$ with length $t-1$.  
    Define $p'(\Q{x}) \doteq \Pr_{D'}(\Q{r}'_t=\Q{x}| \Q{s},\Q{s}')$.
    Then for each $\Q{x}$, $p'(\Q{x}) =p(\pi(\Q{x}))$,
    where $\pi$ is the permutation defined by
    $\pi(\Q{x}) = \Q{x}$ unless $\Q{x}\in \mathcal{X}$ or $f(\Q{x})\in \mathcal{X}$,
    in which case $\pi(\Q{x}) = f(\Q{x})$ and $\pi(f(\Q{x})) = \Q{x}$.
  \end{claim}
  The claim follows by direct calculation based on
  the last line of the experiment.
  \begin{claim}\label{fault}
    Let $\Q{r}$, $\Q{r}'$, $\Q{s}$, and $\Q{s}'$, be as in Claim~\ref{swap}.
    Then
    \[\textstyle \Pr_{D'}[\mbox{\LRU{} faults on $\Q{r}_t$} ~|~ \Q{s},\Q{s}']
    \le \Pr_{D'}[\mbox{$A$ faults on $\Q{r}'_t$} ~|~ \Q{s},\Q{s}'].\]
  \end{claim}
  Why?  It suffices to show that for every item $\Q{x}$ in $\Q{A}$,
  there is a unique item $\Q{y}$ in $\Q{L}$ such that $p'(\Q{x}) \le p(\Q{y})$.
  But by Claim~\ref{swap} and the choice of $\mathcal{X}$,
  this is the case: take $\Q{y}=\Q{x}$ unless $\Q{x}\in \Q{A}-\Q{L}$,
  in which case take $\Q{y}=f(\Q{x}) \in \Q{L}-\Q{A}$.

  Note that in Lemma~\ref{fault}, equality does not necessarily hold because
  $A$ may not have $k$ pages in its cache, or it may have ``irrelevant'' pages
  in its cache --- a page $\Q{x}$ that $D$ requests with {\em less} probability than the
  corresponding page $f(\Q{x})$ in \LRU's cache (so no probability is shifted from 
  $\Q{x}$ to $f(\Q{x})$).

  \begin{claim}\label{cost}
    The first part of the theorem is true: 
    $\E_D[\LRU(\Q{r})] \le \E_{D'}[A(\Q{r}')]$.
  \end{claim}
  This follows directly from Claim~\ref{fault}.
  To see it formally, letting $\Q{s}$ and $\Q{s}'$ range over all 
  equal-length pairs of sequences, we have
  \begin{eqnarray*}
    \E_D[\LRU(\Q{r})]
    &=& \sum_{\Q{s},\Q{s}'}\textstyle
    \Pr_{D'}(\Q{s},\Q{s}')\Pr_{D'}[\mbox{\LRU{} faults on $\Q{r}_{|\Q{s}|+1}$}|\,\Q{s},\Q{s}']
    \\&\le& \sum_{\Q{s},\Q{s}'}\textstyle
    \Pr_{D'}(\Q{s},\Q{s}')\Pr_{D'}[\mbox{$A$ faults on $\Q{r}'_{|\Q{s}'|+1}$}|\,\Q{s},\Q{s}']
    \\&=&\E_{D'}[A(\Q{r}')].
  \end{eqnarray*}
  Above $\Pr_{D'}(\Q{s},\Q{s}')$ denotes the probability that $\Q{s}$ is a prefix of $\Q{r}$
  and $\Q{s}'$ is a prefix of $\Q{r}'$ in the random experiment.

  \begin{claim}\label{cost}
    The second part of the theorem is true: 
    $\E_D[\OPT(\Q{r})] \ge \E_{D'}[\OPT(\Q{r}')]$.
  \end{claim}
  Since the random experiment described above produces the same distribution on
  $\Q{r}$ as $D$ does, it suffices to prove the inequality assuming that the
  pair $(\Q{r},\Q{r}')$ is generated by that experiment.  
  Since $\LRU$ keeps the most recently requested items in its cache,
  and $f:(\Q{A}-\Q{L})\rightarrow(\Q{L}-\Q{A})$,
  we have $\Q{x} \le f(\Q{x})$.
  Thus, in any outcome, $\Q{r}$ dominates $\Q{r}'$ (in the sense of
  Lemma~\ref{domlemma}) and so $\OPT(\Q{r}) \ge \OPT(\Q{r}')$.  This proves the
  claim.
  
  \begin{claim}\label{diffuse}
    The distribution $D'$ defined by the random experiment is in $\Delta_\epsilon$.
  \end{claim}
  This also follows directly from Claim~\ref{swap}.
  To prove it in detail, we need to show that for any $\Q{s}'$ and $\Q{x}$,
  $\Pr_{D'}(\Q{x}|\Q{s}') \le \epsilon$.
  But
  \[\textstyle \Pr_{D'}(\Q{x}|\Q{s}')
  \,=\, \sum_{\Q{s}} \Pr_D(\Q{s})\Pr_{D'}(\Q{r}_t'=\Q{x}|\Q{s},\Q{s}')
  \,\le\, \sum_{\Q{s}} \Pr_D(\Q{s})\, \epsilon
  \,=\, \epsilon.\]
  Above $\Pr_D(\Q{s})$ denotes the probability that $\Q{s}$ is a prefix of $\Q{r}$,
  and $\Q{s}$ ranges over all sequences of length $|\Q{s}'|=t-1$.
  The second-to-last inequality follows because by
  Claim~\ref{swap} each $\Pr_{D'}(\Q{r}_t'=\Q{x}|\Q{s},\Q{s}')$
  equals $\Pr_{D}(\Q{y}|\Q{s})$ for some $\Q{y}$,
  and by the assumption that $D\in\Delta_\epsilon$, $\Pr_{D}(\Q{y}|\Q{s}) \le \epsilon$.
  This proves the claim (and the theorem!).
\end{proof}

\section*{Acknowledgements}
Thanks to Elias Koutsoupias and Lenny Ng for helpful discussions and suggestions.
Thanks to the anonymous referees for helping to clarify the presentation.
\small
\bibliographystyle{plain}
\bibliography{full,competitive}
\end{document}